\documentclass{PoS}
\usepackage{epsfig}
\title{Upper bound on the cutoff in the Standard Model}

\ShortTitle{Upper bound on the cutoff in the Standard Model}

\author{\speaker{M.A.Zubkov}\\
        ITEP, B.Cheremushkinskaya 25, Moscow, 117259, Russia\\
        E-mail: \email{zubkov@itep.ru}}

\author{A.I.Veselov \\
        ITEP, B.Cheremushkinskaya 25, Moscow, 117259, Russia\\
        E-mail: \email{veselov@itep.ru}}

\abstract{The main objective of this presentation is to point out that the
Upper bound on the cutoff in lattice Electroweak theory is still unknown. The
consideration of the continuum theory is based on the perturbation expansion
around trivial vacuum. The internal structure of the lattice Weinberg - Salam
model may appear to be more complicated especially in the region of the phase
diagram close to the phase transition between the physical Higgs phase and the
unphysical symmetric phase of the lattice model, where the continuum physics is
to be approached. We represent the results of our numerical investigation of
the quenched model at infinite bare scalar self coupling $\lambda$. These
results demonstrate that at $\lambda = \infty$ the upper bound on the cutoff is
around $\frac{\pi}{a} = 1.4$ Tev. The preliminary results for finite $\lambda$
are also presented. Basing on these results we cannot yet make a definite
conclusion on the maximal value of the cutoff admitted in the lattice model,
although we have found that the cutoff cannot exceed the value around $1.4 \pm
0.2$ Tev for a certain particular choice of the couplings ($\lambda = 0.009$,
$\beta = 12$, $\theta_W = 30^o$) for the lattices of sizes up to
$12^3\times16$. We also observe that the topological defects, which are to be
identified with quantum Nambu monopoles, dominate in vacuum in the vicinity of
the transition. This indicates that the vacuum of the model is different from
the trivial one. In addition we remind the results of the previous numerical
investigations of the $SU(2)$ gauge - Higgs model, where the maximal reported
value of the cutoff was around $1.5$ Tev.  }

\FullConference{The XXVII International Symposium on Lattice Field Theory - LAT2009\\
         July 26-31 2009\\
         Peking University, Beijing, China}

\begin{document}

\section{Introduction}

 According to
the conventional point of view the upper bound $\Lambda$ on the cutoff in the
Electroweak theory (without fermions) depends on the Higgs mass. It is
decreased when the Higgs mass is increased. And at the Higgs mass around $1$
Tev $\Lambda$ becomes of the order of $M_H$. At the same time for $M_H \sim
200$ Gev the value of $\Lambda$ can be made almost infinite\footnote{Here we do
not consider vacuum stability bound on the Higgs mass related to the fermion
loops.}. This conclusion is made basing on the perturbation expansion around
trivial vacuum. In our presentation we demonstrate that the vacuum of the
lattice Weinberg - Salam model is rather complicated, which means that the
application of the perturbation expansion around trivial vacuum may be limited.

Namely, we investigate the behavior of the topological defects composed of the
lattice gauge fields that are to be identified with quantum Nambu monopoles
\cite{Nambu,BVZ,Chernodub_Nambu}. We show that their lattice density increases
along the lines of constant physics when the ultraviolet cutoff in increased.
At sufficiently large values of the cutoff these objects begin to dominate.

Moving further along the line of constant physics we reach the point on the
phase diagram where the monopole worldlines begin to percolate. This point
roughly coincides with the position of the transition between the physical
Higgs phase and the unphysical symmetric phase of the lattice model. At
infinite bare scalar self coupling $\lambda$ the transition is a crossover and
the ultraviolet cutoff achieves its maximal value around $1.4$ Tev at the
transition  point. At smaller bare values of $\lambda$ correspondent to small
Higgs masses the phase transition becomes stronger. Still we do not know the
order of the phase transition at small values of $\lambda$. We have estimated
the maximal value of the cutoff in the vicinity of the transition point at
$\lambda = 0.009$. The obtained value of the cutoff appears to be around $1.4$
Tev.

\section{The lattice model under investigation}
The lattice Weinberg - Salam Model without fermions contains  gauge field
${\cal U} = (U, \theta)$ (where $ \quad U
 \in SU(2), \quad e^{i\theta} \in U(1)$ are
realized as link variables), and the scalar doublet $ \Phi_{\alpha}, \;(\alpha
= 1,2)$ defined on sites.

The  action is taken in the form
\begin{eqnarray}
 S & = & \beta \!\! \sum_{\rm plaquettes}\!\!
 ((1-\mbox{${\small \frac{1}{2}}$} \, {\rm Tr}\, U_p )
 + \frac{1}{{\rm tg}^2 \theta_W} (1-\cos \theta_p))+\nonumber\\
 && - \gamma \sum_{xy} Re(\Phi^+U_{xy} e^{i\theta_{xy}}\Phi) + \sum_x (|\Phi_x|^2 +
 \lambda(|\Phi_x|^2-1)^2), \label{S}
\end{eqnarray}
where the plaquette variables are defined as $U_p = U_{xy} U_{yz} U_{wz}^*
U_{xw}^*$, and $\theta_p = \theta_{xy} + \theta_{yz} - \theta_{wz} -
\theta_{xw}$ for the plaquette composed of the vertices $x,y,z,w$. Here
$\lambda$ is the scalar self coupling, and $\gamma = 2\kappa$, where $\kappa$
corresponds to the constant used in the investigations of the $SU(2)$ gauge
Higgs model. $\theta_W$ is the Weinberg angle. Bare fine structure
 constant $\alpha$ is expressed through $\beta$ and $\theta_W$ as $\alpha = \frac{{\rm tg}^2 \theta_W}{\pi \beta(1+{\rm tg}^2
\theta_W)}$. In our investigation we fix bare  Weinberg angle equal to $30^o$.
The renormalized fine structure constant can be extracted through the potential
for the infinitely heavy external charged particles.

\section{Phase diagram}

\begin{figure}
\begin{center}
 \epsfig{figure=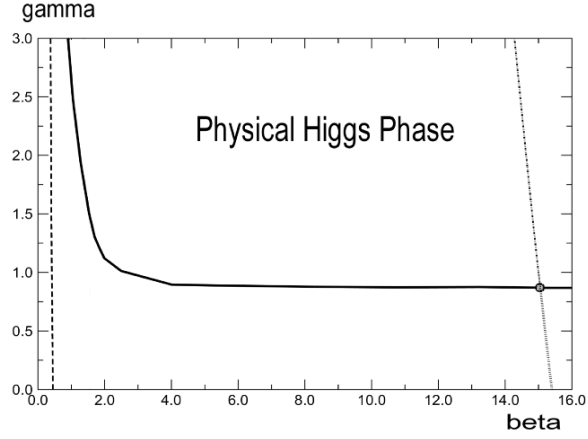,height=60mm,width=80mm,angle=0}
\caption{\label{fig.1} The phase diagram of the model in the
 $(\beta, \gamma)$-plane at infinite $\lambda$.}
\end{center}
\end{figure}
 The phase diagram at infinite $\lambda$ is represented on Fig.1. The dashed vertical line represents the
confinement-deconfinement phase transition corresponding to the $U(1)$
constituent of the model.  The  continuous horizontal line corresponds to the
transition between the broken and the symmetric phases. Real physics is
commonly believed to be achieved within the phase of the model situated in the
right upper corner of Fig.~$1$. The double-dotted-dashed vertical line on the
right-hand side of the diagram represents the line, where the renormalized
$\alpha$ is constant and is equal to $1/128$.

Qualitatively the phase diagram at finite $\lambda$ looks similar to that of
infinite $\lambda$. In the three - dimensional ($\beta, \gamma, \lambda$) phase
diagram the transition surfaces are two - dimensional. The lines of constant
physics on the tree level are the lines ($\frac{\lambda}{\gamma^2} = \frac{1}{8
\beta} \frac{M^2_H}{M^2_W} = {\rm const}$; $\beta = \frac{1}{4\pi \alpha}={\rm
const}$).  In general the cutoff is increased along the line of constant
physics when $\gamma$ is decreased. The maximal value of the cutoff is achieved
at the transition point. Nambu monopole density in lattice units is also
increased when the ultraviolet cutoff is increased.

At $\beta = 12$ the phase diagram is represented on Fig. 2. The physical Higgs
phase is situated up to the transition line. The position of the transition is
localized at the point where the susceptibility extracted from the Higgs field
creation operator achieves its maximum.
\begin{figure}
\begin{center}
 \epsfig{figure=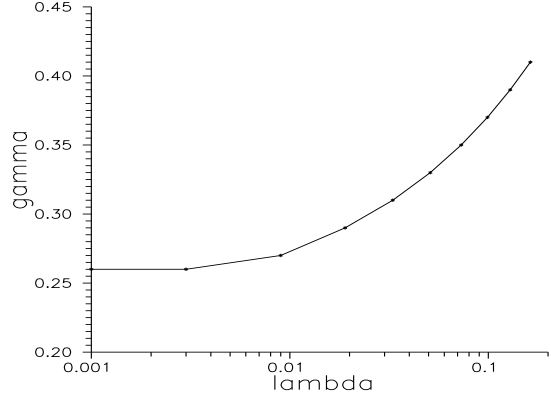,height=60mm,width=80mm,angle=0}
\caption{\label{fig.2} The phase diagram of the model in the
 $(\gamma, \lambda)$-plane at $\beta = 12$.}
\end{center}
\end{figure}

All simulations were performed on lattices of sizes $8^3\times 16$. Several
points were checked using larger lattices up to $16^3\times 24$. At $\lambda =
\infty$ we found no significant difference between the results obtained using
the mentioned lattices. For small $\lambda$ the careful investigation of the
dependence of physical observables on the lattice size has not been performed.

\section{Calculation of the cutoff}

The following variable is considered as creating the $Z$ boson: $ Z_{xy} =
Z^{\mu}_{x} \;
 = {\rm sin} \,[{\rm Arg} (\Phi_x^+U_{xy} e^{i\theta_{xy}}\Phi_y) ]$.
In order to evaluate the masses of the $Z$-boson and the Higgs boson we use the
correlators:
\begin{equation}
\frac{1}{N^6} \sum_{\bar{x},\bar{y}} \langle \sum_{\mu} Z^{\mu}_{x} Z^{\mu}_{y}
\rangle   \sim
  e^{-M_{Z}|x_0-y_0|}+ e^{-M_{Z}(L - |x_0-y_0|)}
\label{corZ}
\end{equation}
and
\begin{equation}
  \frac{1}{N^6}\sum_{\bar{x},\bar{y}}(\langle H_{x} H_{y}\rangle - \langle H\rangle^2)
   \sim
  e^{-M_{H}|x_0-y_0|}+ e^{-M_{H}(L - |x_0-y_0|)},
\label{cor}
\end{equation}
 Here the summation $\sum_{\bar{x},\bar{y}}$ is over the three ``space"
components of the four - vectors $x$ and $y$ while $x_0, y_0$ denote their
``time" components. $N$ is the lattice length in "space" direction. $L$ is the
lattice length in the "time" direction.

In lattice calculations we used two different operators that create Higgs
bosons: $ H_x = |\Phi|$ and $H_x = \sum_{y} Z^2_{xy}$. In both cases $H_x$ is
defined at the site $x$, the sum $\sum_y$ is over its neighboring sites $y$.

After fixing the unitary gauge, lattice Electroweak theory becomes a lattice
$U(1)$ gauge theory. The $U(1)$ gauge field is $ A_{xy}  =  A^{\mu}_{x} \;
 = \,[-{\rm Arg} (\Phi_x^+U_{xy} e^{i\theta_{xy}}\Phi_y)  + 2\theta_{xy}]  \,{\rm mod} \,2\pi$.
The usual Electromagnetic field is $ A_{\rm EM}  =  A + Z^{\prime} - 2 \,{\rm
sin}^2\, \theta_W Z^{\prime}$,
where $Z^{\prime} = [ {\rm Arg} (\Phi_x^+U_{xy} e^{i\theta_{xy}}\Phi_y) ]{\rm
mod} 2\pi$.

The physical scale is given in our lattice theory by the value of the $Z$-boson
mass $M^{phys}_Z \sim 91$ GeV. Therefore the lattice spacing is evaluated to be
$a \sim [91 {\rm GeV}]^{-1} M_Z$, where $M_Z$ is the $Z$ boson mass in lattice
units.

At infinite $\lambda$ the real continuum physics should be approached along the
the line of constant $\alpha_R = \frac{1}{128}$. The  ultraviolet cutoff is
$\Lambda = \frac{\pi}{a} = (\pi \times 91~{\rm GeV})/M_Z$. $\Lambda$ is
increased slowly along this line with decreasing $\gamma$ and achieves the
value around $1.35$ TeV at the transition point between the physical Higgs
phase and the symmetric phase. According to our results this value does not
depend on the lattice size.

In the region of the phase diagram represented on Fig.2 the situation is
similar. Our data obtained on the lattice $8^3\times16$ shows that $\Lambda$ is
increased slowly with the decrease of $\gamma$ at any fixed $\lambda$. We
investigated carefully the vicinity of the transition point at fixed $\lambda =
0.009$ and $\beta = 12$. It has been found that at the transition point the
value of $\Lambda$ is equal to $1.4 \pm 0.2$ Tev. The first check of a larger
lattice (of size $12^3\times 16$) does not show an increase of this value.
 However, the careful
investigation of the dependence of $\Lambda$ on the lattice size (as well as on
$\lambda$) is to be the subject of future investigations.

On Fig. 3 the dependence of $M_Z$ in lattice units on $\gamma$ is represented
at $\lambda =0.009$ and $\beta = 12$, where $\gamma_c = 0.273 \pm 0.002$. .
Unfortunately, we cannot yet estimate the renormalized Higgs boson mass due to
the lack of statistics. However, we expect it does not deviate significantly
from the tree level estimate $M^0_H = \frac{\sqrt{8\beta \lambda}}{\gamma}
\times 80$ Gev. In the vicinity of the phase transition at $\lambda =0.009$,
$\beta = 12$ bare value of the Higgs mass is $M^0_H \sim 270$ Gev.

\begin{figure}
\begin{center}
 \epsfig{figure=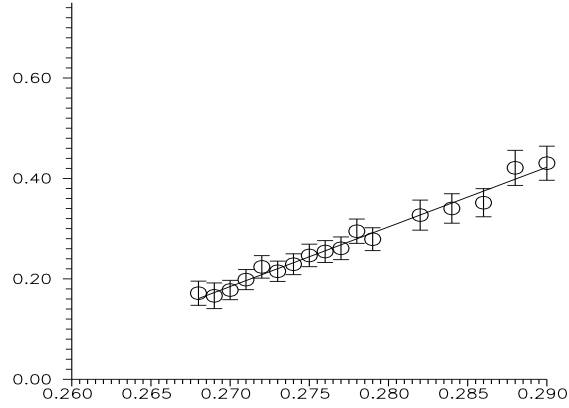,height=60mm,width=80mm,angle=0}
\caption{\label{fig.3} Z - boson mass in lattice units at $\lambda =0.009$ and
$\beta = 12$. }
\end{center}
\end{figure}

\section{The renormalized coupling}

The bare constant $\alpha = e^2/4\pi$ (where $e$ is the electric charge) can be
easily calculated in our lattice model. It is found to be equal to $1/(4\pi
\beta)$. Therefore, its physical value $\alpha(M_Z)\sim 1/128$ could be
achieved at the values of $\beta$ in some vicinity of $10$. This naive guess
is, however, to be corrected by the calculation of the renormalized coupling
constant $\alpha_R$. We perform this calculation using the potential for
infinitely heavy external fermions. We consider Wilson loops for the
right-handed external leptons: $
 {\cal W}^{\rm R}_{\rm lept}(l)  =
 \langle {\rm Re} \,\Pi_{(xy) \in l} e^{2i\theta_{xy}}\rangle.
$
Here $l$ denotes a closed contour on the lattice. We consider the following
quantity constructed from the rectangular Wilson loop of size $r\times t$:
$
 {\cal V}(r) = \lim_{t \rightarrow \infty}{
 \rm log}
 \frac{  {\cal W}(r\times t)}{{\cal W}(r\times (t+1))}.
$
At large enough distances we expect the appearance of the Coulomb interaction
$
 {\cal V}(r) = -\frac{\alpha_R}{r} + const.
$

The renormalized coupling constant $\alpha$ is found to be close to the
realistic value $\alpha(M_Z)=1/128$ along the line represented in Fig.~$1$ (at
$\lambda \rightarrow \infty$) in the vicinity of $\beta = 15$. We do not
observe any dependence of $\alpha_R$ on the lattice size at $\lambda = \infty$.

 At $\lambda = 0.009$, $\beta = 12$, $ \gamma = \gamma_c(\lambda)\sim 0.273$ the renormalized fine
structure constant calculated on the lattice $8^3\times 16$ is $\alpha_R =
\frac{1}{98\pm 3}$. The same value has been obtained also on the larger lattice
($12^3\times 16$), which shows that the value of $\alpha_R$ does not depend on
the lattice size also for the small values of $\lambda$.

\section{Nambu monopole density }

\begin{figure}
\begin{center}
 \epsfig{figure=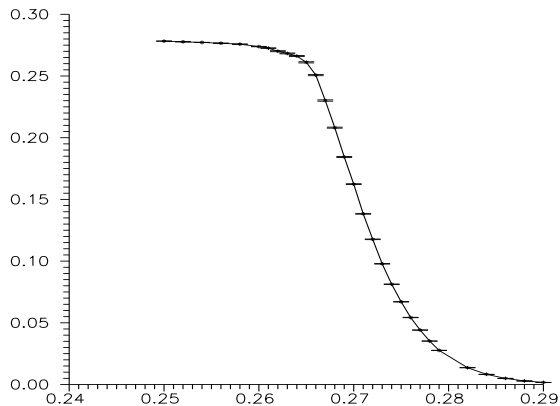,height=60mm,width=80mm,angle=0}
\caption{\label{fig.4} Nambu monopole density  as a function of $\gamma$ at
$\lambda = 0.009$, $\beta = 12$.  }
\end{center}
\end{figure}

According to \cite{BVZ,Chernodub_Nambu,VZ2008} the worldlines of the quantum
Nambu monopoles can be extracted from the field configurations as follows:
\begin{equation}
 j_A =  \frac{1}{2\pi} {}^*d([d A]{\rm mod}2\pi)
\end{equation}
(The notations of differential forms on the lattice \cite{forms} are used
here.) The monopole density is defined as $
 \rho = \left\langle \frac{\sum_{\rm links}|j_{\rm link}|}{4L^4}
 \right\rangle,
$
where $L$ is the lattice size.

In Fig.~$4$ we represent Nambu monopole density as a function of $\gamma$ at
$\lambda = 0.009$, $\beta = 12$. The point of the transition is localized as
the position of the maximum of the susceptibility  $\chi = \langle H^2 \rangle
- \langle H\rangle^2$ extracted from $H = \sum_{y} Z^2_{xy}$. The value of
monopole density at $\gamma_c = 0.273$, $\beta = 12$, $\lambda = 0.009$ is
around $0.1$. At this point the value of the cutoff is $\Lambda \sim 1.4 \pm
0.2$ Tev. The monopole density around $0.1$ means that among $10$ sites there
exist $4$ sites that are occupied by the monopole. Average distance between the
two monopoles is, therefore, less than $1$ lattice spacing and it is not
possible at all to speak of the given configurations as of representing the
single Nambu monopole.

That's why these complicated configurations constructed of the gauge field and
the scalar field dominate in vacuum in the vicinity of the transition point.
This means that the usual perturbation expansion around trivial vacuum (gauge
field equal to zero) may not be valid in a vicinity of the phase transition
between the physical Higgs phase and the unphysical symmetric phase of the
model. This might explain why we do not observe in our numerical simulations
the large values of $\Lambda$ predicted by the conventional perturbation
theory.

\section{Conclusions}

In Table $1$ we list the values of the lattice spacing used in selected lattice
studies of $SU(2)$ Gauge - Higgs Model. From this table it is clear that the
correspondent value of the cutoff $\frac{\pi}{a}$ does not exceed $1.5$ Tev.

\begin{table}
\label{tab.01}
\begin{center}
\begin{tabular}{|c|c|c|}
\hline
{\bf Reference}  & {\bf inverse lattice spacing} $\frac{1}{a}$ (GeV) & {\bf $M_H$} (GeV)\\
\hline
\cite{1}  & 140 (space direction) 570 (time direction) & 80 \\
\hline \cite{2}  & 280 (time direction) & 80 \\
\hline \cite{3}  & 280 & 34 \\
\hline \cite{4}  & 110 & 16 \\
\hline \cite{5}  & 90 (space direction) 350 (time direction) & 34 \\
\hline \cite{6}  & 280 & 48 \\
\hline \cite{7}  & 140 & 35 \\
\hline \cite{8}  & 280 & 20 , 50 \\
\hline \cite{9}  & 190 & 50 \\
\hline \cite{10}  & 260 & 57 - 85 \\
\hline \cite{11}  & 200 - 300 & 47 - 108 \\
\hline \cite{12}  & 400 & 480 \\
\hline \cite{13}  & 330 -  470 & 280 - 720
 \\
\hline \cite{14}  & 250 -  470 &  720
($\lambda =\infty$) \\
\hline
\end{tabular}
\end{center}
\end{table}

Our own numerical data demonstrate that the vacuum structure of the lattice
Weinberg - Salam model is rather complicated. Namely, the topological defects
identified with quantum Nambu monopoles dominate in vacuum in the vicinity of
the phase transition between the symmetric phase and the Higgs phase. This
indicates that the usual perturbation expansion around trivial vacuum may not
be applied in this region of the phase diagram. As a consequence one cannot
apply the conventional perturbation theory to the evaluation of the Ultraviolet
cutoff upper bound in this region of the phase diagram. Qualitatively this
situation seem to us similar to that of the Ginzburg - Landau theory of
superconductivity. Within this theory in a certain vicinity of the phase
transition the fluctuations of the order parameter become so strong that the
perturbation expansion around the trivial solution of Ginzburg - Landau
equations cannot be applied.

Thus we conclude that the upper bound on the Ultraviolet cutoff in the lattice
Electroweak theory is still not known. The establishing of this upper bound is
to be a subject of future investigations. We suppose that the upper bound on
the cutoff obtained with the aid of nonperturbative lattice methods may differ
from the conventional one obtained via the perturbation expansion around
trivial vacuum.

\begin{acknowledgments}

This work was partly supported by RFBR grants 09-02-08308, 09-02-00338,
08-02-00661, and 07-02-00237, by Grant for leading scientific schools
679.2008.2,by Federal Program of the Russian Ministry of Industry, Science and
Technology No 40.052.1.1.1112. The numerical simulations have been performed
using the facilities of Moscow Joint Supercomputer Center.

\end{acknowledgments}

\end{document}